\documentstyle[12pt]{article}
\begin{document}
\begin{center}
{\Large \bf Mass hierarchy of leptons and hadrons within
the framework of electrodynamics}
\bigskip

{\large D.L.~Khokhlov}
\smallskip

{\it Sumy State University, R.-Korsakov St. 2\\
Sumy 244007 Ukraine\\
e-mail: khokhlov@cafe.sumy.ua}
\end{center}

\begin{abstract}
Structure of leptons and hadrons within
the framework of electrodynamics is considered.
Muon and tau-lepton have the structure of 3 electrons.
The mass of muon is defined by the section
of two-photon annihilation.
The mass of tau-lepton is defined by the section
of three-photon annihilation.
Hadrons are characterized by the structure
of 5 electrons. The masses of hadrons are defined via
the masses of muon and tau-lepton.
\end{abstract}

According to the standard theory, hadrons have the quark structure.
Baryons consist of 3 quarks, mesons consist of the quark-antiquark
pair.
In the $SU(2)\times U(1)$ electroweak theory~\cite{B},
masses of the fermions, leptons and quarks, arise due to the
Yukawa interaction
\begin{equation}
f(\bar\psi_{L}\psi_{R}\varphi+\bar\psi_{R}\psi_{L}\bar\varphi)
\label{eq:yui}
\end{equation}
where the couplings $f$ are different for each fermion.
The standard theory do not define the couplings $f$
and cannot explain the hierarchy of the leptons and
quarks masses.

To explain the mass hierarchy between lepton
generations within the framework of electrodynamics
it was proposed~\cite{Kh1} to consider
muon and tau-lepton as composite particles.
It was assumed that muon and tau-lepton have the following structure
\begin{equation}
\mu^-\equiv e^-e^+e^-\label{eq:a}
\end{equation}
\begin{equation}
\tau^-\equiv e^-e^+e^-.\label{eq:b}
\end{equation}
Muon arises due to the reaction
\begin{equation}
e^- +2\gamma\rightarrow e^-e^+e^-,\label{eq:c}
\end{equation}
and tau-lepton arises due to the reaction
\begin{equation}
e^- +3\gamma\rightarrow e^-e^+e^-.\label{eq:d}
\end{equation}
The mass of muon is defined by the cross section
of two-photon annihilation
\begin{equation}
m_{\mu}=\frac{\hbar}{r_{2\gamma}}=70\ {\rm MeV}.
\label{eq:mmu}
\end{equation}
The experimental value is $m_{\mu}=106\ {\rm MeV}$~\cite{R}.
The mass of tau-lepton is defined by the cross section
of three-photon annihilation
\begin{equation}
m_{\tau}=\frac{\hbar}{r_{3\gamma}}=2200\ {\rm MeV}.
\label{eq:mta}
\end{equation}
The experimental value is $m_{\tau}=1784\ {\rm MeV}$~\cite{R}.

In order to describe the decays of muon
and tau-lepton within the framework of electrodynamics
the following reactions are introduced
\begin{equation}
\gamma\rightarrow\bar\nu_{e}\bar\nu_{e}
\label{eq:gnu}
\end{equation}
\begin{equation}
\bar\nu_{e}+\gamma\rightarrow\nu_{\mu}
\label{eq:x}
\end{equation}
\begin{equation}
\bar\nu_{e}+2\gamma\rightarrow\nu_{\tau}.
\label{eq:y}
\end{equation}
Combining eqs.~(\ref{eq:a}), (\ref{eq:c}),
(\ref{eq:gnu}), (\ref{eq:x})
we obtain the reaction
for the decay of muon
\begin{equation}
\mu^-\equiv e^-e^+e^-\rightarrow e^- +2\gamma
\rightarrow e^- +\gamma\bar\nu_{e}\bar\nu_{e}
\rightarrow e^- +\bar\nu_{e}\nu_{\mu}\label{eq:dm}.
\end{equation}
Combining eqs.~(\ref{eq:b}), (\ref{eq:d}),
(\ref{eq:gnu}), (\ref{eq:y})
we obtain the reaction
for the decay of tau-lepton
\begin{equation}
\tau^-\equiv e^-e^+e^-\rightarrow e^- +3\gamma
\rightarrow e^- +2\gamma\bar\nu_{e}\bar\nu_{e}
\rightarrow e^- +\bar\nu_{e}\nu_{\tau}\label{eq:dt}.
\end{equation}

Let us assume that hadrons are characterized by the structure
of 5 electrons
\begin{equation}
e^-e^+e^-e^+e^-.
\label{eq:had}
\end{equation}
In addition to reactions (\ref{eq:gnu}), (\ref{eq:x}), (\ref{eq:y}),
let us introduce the following reactions
\begin{equation}
\bar\nu_{e}+3\gamma\rightarrow\nu_{e}
\label{eq:bee}
\end{equation}
\begin{equation}
\bar\nu_{e}+4\gamma\rightarrow\bar\nu_{e}.
\label{eq:ee}
\end{equation}

The structure of the meson-antimeson pair consists of ten
electron-positron pairs. Let us assume that
three electron-positron pairs transform into muon or tau-lepton pairs
\begin{equation}
e^-e^+e^-e^+e^-e^+\rightarrow n(l^- l^+)
\label{eq:me3}
\end{equation}
where $l$ denotes muon or tau-lepton, $n$ is the number of pairs
equal to 1 or 3.
Two electron-positron pairs transform into the pair of
neutrino-antineutrino
\begin{equation}
e^-e^+e^-e^+\rightarrow 4\gamma \rightarrow
\bar\nu\bar\nu+3\gamma\rightarrow\bar\nu\nu.
\label{eq:me2}
\end{equation}

Consider the structure of pion which within the framework of
the standard theory consists of $u, d$ quarks.
Pair of the charged pions has the structure
\begin{equation}
\pi^-\pi^+\equiv e^-e^+e^-e^+e^- + e^+e^-e^+e^-e^+
\rightarrow \mu^- \mu^+ +\bar\nu_{\mu}\nu_{\mu}.
\label{eq:pi}
\end{equation}
The mass of pion is estimated as
\begin{equation}
m_{\pi}=m_{\mu}=106\ {\rm MeV}.
\label{eq:mpi}
\end{equation}
The experimental value is $m_{\pi}=140\ {\rm MeV}$~\cite{R}.
The pion structure of 5 electrons allows one to explain
the probability of electron-positron annihilation into pions
\begin{equation}
\frac{\Gamma(e^-e^+\rightarrow \pi\pi)}
{\Gamma(e^-e^+\rightarrow \mu\mu)}=\frac{\sum q^{2}_i(\pi)}
{\sum q^{2}_i(\mu)}=\frac{5}{3}.
\label{eq:GG}
\end{equation}

Consider the structure of $K$ meson which within the framework of
the standard theory includes strange $s$ quark.
Pair of the charged $K$ mesons has the structure
\begin{equation}
K^-K^+\equiv e^-e^+e^-e^+e^- + e^+e^-e^+e^-e^+
\rightarrow 3\mu^- 3\mu^+ +\bar\nu_{\mu}\nu_{\mu}.
\label{eq:Ka}
\end{equation}
The mass of $K$ meson is estimated as
\begin{equation}
m_{K}=3m_{\mu}=3 \times 106=318\ {\rm MeV}.
\label{eq:mK}
\end{equation}
The experimental value is $m_{K}=494\ {\rm MeV}$~\cite{R}.

Consider the structure of $D$ meson which within the framework of
the standard theory includes charm $c$ quark.
Pair of the charged $D$ mesons has the structure
\begin{equation}
D^-D^+\equiv e^-e^+e^-e^+e^- + e^+e^-e^+e^-e^+
\rightarrow \tau^- \tau^+ +\bar\nu_{\tau}\nu_{\tau}.
\label{eq:D}
\end{equation}
The mass of $D$ meson is estimated as
\begin{equation}
m_{D}=m_{\tau}=1784\ {\rm MeV}.
\label{eq:mD}
\end{equation}
The experimental value is $m_{D}=1869\ {\rm MeV}$~\cite{R}.

Consider the structure of $B$ meson which within the framework of
the standard theory includes beauty $b$ quark.
Pair of the charged $B$ mesons has the structure
\begin{equation}
B^-B^+\equiv e^-e^+e^-e^+e^- + e^+e^-e^+e^-e^+
\rightarrow 3\tau^- 3\tau^+ +\bar\nu_{\tau}\nu_{\tau}.
\label{eq:B}
\end{equation}
The mass of $B$ meson is estimated as
\begin{equation}
m_{B}=3 \times m_{\tau}=3 \times 1784=5352\ {\rm MeV}.
\label{eq:mB}
\end{equation}
The experimental value is $m_{B}=5271\ {\rm MeV}$~\cite{R}.

Consider the structure of proton which within the framework of
the standard theory consists of $u, u, d$ quarks.
Proton structure can be given by
\begin{equation}
p\equiv e^+ +2\gamma+3\gamma\rightarrow e^+ + e^-e^+ + e^-e^+
\label{eq:pro}
\end{equation}
where electron-positron pairs are in the superposition of
two-photon annihilation state and three-photon annihilation state.
The mass of proton is estimated as
\begin{equation}
m_{p}=\frac{1}{2} (m_{\mu} + m_{\tau})=\frac{106 + 1784}{2}=945\ {\rm MeV}.
\label{eq:mp}
\end{equation}
The experimental value is $m_{p}=938\ {\rm MeV}$~\cite{R}.
In view of (\ref{eq:gnu}), (\ref{eq:ee}), (\ref{eq:pro}), 
the decay of proton is given by
\begin{equation}
p\equiv e^+e^-e^+e^-e^+\rightarrow e^+ +5\gamma
\rightarrow e^+ +4\gamma + \nu_{e}\nu_{e}
\rightarrow e^+ +\nu_{e}\nu_{e}.
\label{eq:dp}
\end{equation}
The birth of the pair of identical neutrinos violates
the particle-antiparticle conservation law, so
the decay of proton with the structure given by eq.~(\ref{eq:pro})
is forbidden.

\end{document}